\documentclass[a4paper,oneside,draft]{article}
\usepackage{amsmath}

\input{tcilatex}

\begin{document}

\title{Teleporting a rotation on remote photons}
\author{Guo-Yong Xiang, Jian Li, Guang-Can Guo \\
Key Laboratory of Quantum Information, and Department of Physics,\\
University of Science and Technology of China, Hefei 230026 China}
\maketitle

\begin{abstract}
Quamtum remote rotation allows implement local quantum operation on remote
systems with shared entanglement. Here we report an experimental
demonstration of remote rotation on single photons using linear optical
element. And the local dephase is also teleported during the process. The
scheme can be generalized to any controlled rotaion commutes with $\sigma
_{z}$.

PACS: 03.67.Lx, 03.67.Hk, 03.65.Od, 42.50.-P
\end{abstract}

The development of quantum information in the last two decades promises the
powerful applications in information manipulation\cite{NC}. Using quantum
operations, some missions, for example large number factoring, can be solved
by quantum algorithm effectively, which is impossible for classical
computer. Arbitrary rotation gates on single qubit and controlled-NOT(CNOT)
gate on two qubits are sufficient and necessary, for universal quantum
computation\cite{RZBB}. These two kinds of operations have been realized
locally in laboratory on physical systems such as trapped ion, neutral atom,
and so on\cite{NC}. If the qubits involved in the process are separated in
distance(e.g. distributed computation\cite{G}\cite{CEHM}), operations must
be implemented nonlocally via sharing entanglement, local operation and
classical communication, i.e. remote operation\cite{EJPP}\cite{HVCP}\cite%
{CLP}\cite{RAG}\cite{HPV}\cite{HRYDG}.

Suppose only two parties, Alice and Bob, are involved in the process\cite%
{HPV}. The former has the operation gate, while the later has the state to
be operated on. The trivial realisation of remote operation can be finished
by the bi-directional quantum state teleportation. The target state is
teleported from Bob to Alice. Then Alice applies the operation on the state
and sends the resulting state back to Bob via another state teleportation.
The total resources for this trivial protocol are two maximally-entangled
states (2 ebits) shared and two bits of classical communication (2 cbits) in
each direction. And there's no restriction for either the target state or
the operation. If the gate is restricted in the unitary operation set $%
\mathsf{U}_{com}\cup \mathsf{U}_{anti}$, where $\mathsf{U}_{com}$ ($\mathsf{U%
}_{anti}$) is the set of single qubit operation commuting (anti-commuting)
with $\sigma _{z}$, the resource can be reduced to two ebits shared, two
cbits from Bob to Alice and one cbit reversely in the optimal nontrivial
scheme. Moreover, if the set is either $\mathsf{U}_{com}$ or $\mathsf{U}%
_{anti}$, the scheme can be simplified by using only one ebit shared and one
cbit in each direction\cite{HPV}. Specially, if the operation at Alice's
site is a collective one with an auxiliary qubit, the two-qubit gate can be
performed nonlocally\cite{EJPP}\cite{HRYDG}.

Here we present an experimental demonstration of a remote rotation on single
photons. Unitary rotation commuting with $\sigma _{z}$ is implemented
remotely on the polarization qubit that Bob hold, using the entanglement
between the polarization qubit of one photon and the path qubit of another
one from spontaneous parametric down conversion (SPDC).

Let's consider the remote rotation on single qubit. Alice has an operation
gate $U_{com}=e^{i\varphi \sigma _{z}/2}$ to implement and Bob has an
arbitrary state $\left\vert \psi \right\rangle _{3}=\left\vert \psi (\theta
,\phi )\right\rangle _{3}=\alpha \left\vert 0\right\rangle _{3}+\beta
\left\vert 1\right\rangle _{3}$ to be operated on, where $\alpha =\cos
\theta $, $\beta =e^{i\phi }\sin \theta $ and $\left\{ \left\vert
0\right\rangle ,\left\vert 1\right\rangle \right\} $ are the basis of $%
\sigma _{z}$. Geometrically, the operation rotates the state vector in the
Bloch sphere through angle $\varphi $ about $z$ axis,$\left\vert \psi
(\theta ,\phi )\right\rangle _{3}\rightarrow \left\vert \psi (\theta ,\phi
^{\prime }=\phi +\varphi )\right\rangle _{3}$ (see Fig.1a). That is the
teleportation of an angle. Specially, the gate is an identity operation for $%
\varphi =0$, while $\sigma _{z}$ for $\varphi =\pi $. Given the shared
entanglement $\left\vert \Phi ^{+}\right\rangle _{12}=\frac{1}{\sqrt{2}}%
\left( \left\vert 0\right\rangle _{1}\left\vert 0\right\rangle
_{2}+\left\vert 1\right\rangle _{1}\left\vert 1\right\rangle _{2}\right) $,
the scheme can be realized by three steps (see Fig.1b).

\textit{i}) \textit{Encoding}: Bob performs a CNOT operation on his qubits, 
\textit{2} and \textit{3}, where \textit{3} is the controller. Then a $%
\sigma _{z}$ measurement on \textit{2} is made by Bob and the result is sent
to Alice via one cbit classical communication. Alice would perform bit-flip
operation $\sigma _{x}$ on her qubit \textit{1} for result $\left\vert
1\right\rangle _{2}$, while doing nothing for $\left\vert 0\right\rangle
_{2} $. The remaining system of qubits \textit{1} and \textit{3}, are in
state 
\begin{equation}
\left\vert \psi \right\rangle _{13}=\alpha \left\vert 0\right\rangle
_{1}\left\vert 0\right\rangle _{3}+\beta \left\vert 1\right\rangle
_{1}\left\vert 1\right\rangle _{3}.
\end{equation}%
Thus the state $\left\vert \psi \right\rangle _{3}$ to be operated on is
encoded into $\left\vert \psi \right\rangle _{13}$ within the Hilbert
subspace $H^{2}=\left\{ \left\vert 0\right\rangle _{1}\left\vert
0\right\rangle _{3},\left\vert 1\right\rangle _{1}\left\vert 1\right\rangle
_{3}\right\} $ of the composite system , \textit{1} and \textit{3}.

\textit{ii}) \textit{Operating}: Alice implements the required quantum gate $%
U_{com}$ on her qubit \textit{1}, while Bob do nothing,%
\begin{equation}
U_{com}\otimes I_{3}\left\vert \psi \right\rangle _{13}=\alpha e^{i\varphi
/2}\left\vert 0\right\rangle _{1}\left\vert 0\right\rangle _{3}+\beta
e^{-i\varphi /2}\left\vert 1\right\rangle _{1}\left\vert 1\right\rangle _{3}.
\end{equation}%
Here the local gate plays the role of global rotaion within $H^{2}$. The
desired state $\left\vert \psi ^{\prime }\right\rangle
_{3}=U_{com}\left\vert \psi \right\rangle _{3}$ has been just embedded in
the composite system.

\textit{iii}) \textit{Decoding}: A $\sigma _{x}$ measurement $\{\left\vert
\pm \right\rangle =\frac{1}{\sqrt{2}}(\left\vert 0\right\rangle \pm
\left\vert 1\right\rangle )\}$ is performed on \textit{1} by Alice and the
result is sent to Bob. Qubit \textit{3} will be found in the desired state $%
\left\vert \psi ^{\prime }\right\rangle _{3}$ for the result $\left\vert
+\right\rangle _{1}$ or $\left\vert \psi ^{\prime }\right\rangle
_{3}=U_{com}\sigma _{z}\left\vert \psi \right\rangle _{3}$ in for $%
\left\vert -\right\rangle _{1}$.\ The latter can be converted into $%
\left\vert \psi ^{\prime }\right\rangle _{3}$ by an additional $\sigma _{z}$
rotation for $\sigma _{z}U_{com}\sigma _{z}=U_{com}$. That is, $\left\vert
\psi ^{\prime }\right\rangle _{3}$ is decoded out from the composite system.

The total resources needed in the whole process are one ebit shared, one
cbit from Bob to Alice for encoding and one cbit from Alice to Bob for
decoding.

To realise the protocol above, the key is how to choose physical qubit,
local operation gate and entangled state. For photons, both the polarization 
$\left\{ \left\vert H\right\rangle ,\left\vert V\right\rangle \right\} $ and
the path $\left\{ \left\vert u\right\rangle ,\left\vert d\right\rangle
\right\} $ can represent the logic states $\{\left\vert 0\right\rangle
,\left\vert 1\right\rangle \}$ for qubits. For polarization qubit, arbitrary
unitary rotation can be performed by using half-wave plate(HWP) and
quarter-wave plate(QWP)\cite{JKMW}. The controlled-NOT gate between
polarization qubit and path qubit of the same photon can be operated on by
the polarization beam splitter(PBS), $\left\vert H\right\rangle \left\vert
u\right\rangle (\left\vert H\right\rangle \left\vert d\right\rangle
)\rightarrow \left\vert H\right\rangle \left\vert u^{\prime }\right\rangle
(\left\vert H\right\rangle \left\vert d^{\prime }\right\rangle )$, $%
\left\vert V\right\rangle \left\vert u\right\rangle (\left\vert
V\right\rangle \left\vert d\right\rangle )\rightarrow \left\vert
H\right\rangle \left\vert d^{\prime }\right\rangle (\left\vert
H\right\rangle \left\vert u^{\prime }\right\rangle )$\cite{CAK}. And the
bi-photon states entangled between polarization or path qubit can be
generated via SPDC process\cite{K}. Here the three qubits are embedded in
the three freedoms of two photons distributed to Alice and Bob,
polarization(qubit \textit{1}) of photon \textit{A}, path(qubit \textit{2})
and polarization(qubit \textit{3}) of photon \textit{B}.

The experimental setup is shown in Fig.2. A mode-locked Ti:Sapphire pulsed
laser (with the pulse width less than 200 fs, the repetition about 82MHz and
the center-wavelength at 780.0nm) is frequency-doubled to produce the
pumping source for SPDC process. A 1.0mm thick BBO crystal cut for type-II
phase match is used as the down converter. By the non-collinear degenerated
SPDC procss, two photons, \textit{A} and\textit{\ B}, are produced in the
polarization-entangled state $\left\vert \Psi ^{+}\right\rangle _{AB}=\frac{1%
}{\sqrt{2}}(\left\vert H\right\rangle _{A}\left\vert V\right\rangle _{B}\pm
\left\vert V\right\rangle _{A}\left\vert H\right\rangle _{B})$\cite{K}. Bob
uses PBS \textit{P1} to split photon \textit{B} in two paths $\left\{
\left\vert u\right\rangle ,\left\vert d\right\rangle \right\} $ and HWP 
\textit{H1} at $45^{\circ }$ as a $\sigma _{x}$ gate\ is used to flip the
polarization in path \textit{u}. Hence the polarization entanglement between
the two photons is converted into polarization-path entanglement,%
\begin{equation}
\left\vert \Psi ^{+}\right\rangle _{123}=\frac{1}{\sqrt{2}}(\left\vert
H\right\rangle _{1}\left\vert u\right\rangle _{2}+\left\vert V\right\rangle
_{1}\left\vert d\right\rangle _{2})\left\vert H\right\rangle _{3}.
\label{ES}
\end{equation}%
The polarization of \textit{B} can be prepared in arbitrary state $%
\left\vert \psi \right\rangle _{3}=\alpha \left\vert H\right\rangle
_{3}+\beta \left\vert V\right\rangle _{3}$ with identical sets of
waveplates, $\{H_{u},Q_{u}\}$ and $\{H_{d},Q_{d}\}$, in each path\cite{JKMW}%
. The global state is initialized in $\left\vert \Phi ^{+}\right\rangle
_{12}\left\vert \psi \right\rangle _{3}=\frac{1}{\sqrt{2}}(\left\vert
H\right\rangle _{1}\left\vert u\right\rangle _{2}+\left\vert V\right\rangle
_{1}\left\vert d\right\rangle _{2})(\alpha \left\vert H\right\rangle
_{3}+\beta \left\vert V\right\rangle _{3}).$ The three steps for remote
rotation are performed as follows.

\textit{i'}) \textit{Encoding}: Path \textit{u} and \textit{d} of photon 
\textit{B} are input in a PBS \textit{P2} to perform a \textit{CNOT}
operation, where polarization is the controlling qubit and path is the
target one. The optical path length of \textit{u} and \textit{p} are tuned
to be equal to ensure no relative phase factor between the two terms in eq.%
\ref{ES}. The $\sigma _{z}$ measurement on qubit \textit{2} is completed by
reading out the path information of photon \textit{B}. If \textit{B} is
found in path \textit{u}', $\left\vert \psi \right\rangle _{B}$ is encoded
into $\left\vert \psi \right\rangle _{AB}=\alpha \left\vert H\right\rangle
_{A}\left\vert H\right\rangle _{B}+\beta \left\vert V\right\rangle
_{A}\left\vert V\right\rangle _{B}$. Or if \textit{B} is found in path 
\textit{d}', the two photons will be in $\left\vert \psi ^{\prime
}\right\rangle _{13}=\alpha \left\vert V\right\rangle _{1}\left\vert
H\right\rangle _{3}+\beta \left\vert H\right\rangle _{1}\left\vert
V\right\rangle _{3}$, which can be transformed into $\left\vert \Psi
\right\rangle _{13}$ by another HWP at $45^{\circ }$ on photon \textit{A}.
Here we omit the later case without loss of generality. The polarization
state of photon \textit{B} is encoded in $\left\{ \left\vert H\right\rangle
_{1}\left\vert H\right\rangle _{3},\left\vert V\right\rangle _{1}\left\vert
V\right\rangle _{3}\right\} $\cite{PJF}.

\textit{ii'}) \textit{Operat}ing: The operation $U_{com}$ can be performed
by a pair of QWP at $45^{\circ }$ with a HWP at $\frac{\varphi }{2}%
-45^{\circ }$ between them. Such device has been used to verify the
geometric phase of classical light and photons\cite{HR}\cite{BDM}. For
single qubit operation, any additional global phase is trivial, so $U_{com}$
can be replaced by $e^{i\varphi /2}U_{com}$, which can be realised by one
zero-order waveplate at $0^{\circ }$ tilted in a suitable angle(see ref.\cite%
{KWWAE} for similar application). Here we chose $\varphi =120^{\circ }$ by a
tilted QWP \textit{Q1}.

iii') Decoding: Alice makes her $\sigma _{x}$ measurement $\{\left\vert
D\right\rangle _{1}=\frac{1}{\sqrt{2}}\left( \left\vert H\right\rangle
_{1}+\left\vert V\right\rangle _{1}\right) ,\left\vert C\right\rangle _{1}=%
\frac{1}{\sqrt{2}}\left( \left\vert H\right\rangle _{1}-\left\vert
V\right\rangle _{1}\right) \}$ using polarizer. Photon \textit{A} is
detected by a single photon detector(SPCM-AQR-14 by EG\&G). Photon \textit{B}
will be collapsed into $\left\vert \psi ^{\prime }\right\rangle
_{3}=U_{com}\left\vert \psi \right\rangle _{3}$ for result $\left\vert
+\right\rangle _{1}$, and $\left\vert \psi ^{\prime \prime }\right\rangle
_{3}=U_{com}\sigma _{z}\left\vert \psi \right\rangle _{3}$ for result $%
\left\vert -\right\rangle _{1}$. The latter can be converted into $%
\left\vert \psi ^{\prime }\right\rangle _{B}$ by a HWP at $0^{\circ }$, i.e.
a $\sigma _{z}$ rotation. The ploarization state of photon \textit{B} is
reconstructed by quantum state tomography using polarization analyzer and
detector. The measurement on \textit{A} and \textit{B} are collected for
coincidence count with the window time 5ns.

In the real experiment, there are two kinds of imperfection which induce the
phase decoherence. One is caused by the birefregency of BBO, which induces
the partial time-seperation between the wavepackets of two polarizations. It
can be described by Kraus operators $\{\sqrt{\frac{1+p}{2}}I,\sqrt{\frac{1-p%
}{2}}\sigma _{z}\}$ on photon \textit{A} or \textit{B}. Here \textit{p} is
just the visibility of the entangled state from SPDC. The other one is the
mismatching of space mode in PBS \textit{P2}(see Fig.3). The two PBSs, 
\textit{P1} amd \textit{P2} at Bob's site consist of a Mach-Zedner
interferometer. The mode-mismatch can be represented by a process dephase $\{%
\sqrt{\frac{1+\eta }{2}}I,\sqrt{\frac{1-\eta }{2}}\sigma _{z}\}$ on the
paths of \textit{B}, where $\eta $ is the visibility of interferometer.
Because the symmetry between two qubit in $\left\vert \Phi ^{+}\right\rangle 
$, both of the imperfections can be considered to be performed on the
polarization of photon \textit{A}. Further, it can also be regarded as a
control phase operation on qubit \textit{1} and an anxiliary system \textit{%
1'}, where the later is the target. Since the control phase gate $%
CP_{1^{\prime }1}=\left\vert 0\right\rangle _{1^{\prime }1^{\prime
}}\left\langle 0\right\vert \otimes I+\left\vert 1\right\rangle _{1^{\prime
}1^{\prime }}\left\langle 1\right\vert \otimes \sigma _{z}$ commuting with $%
\sigma _{z}$ on \textit{1}, the dephasing, or the control phase operation in
the extended systems, is included in $\mathsf{U}_{com}$. That is the
nonlocal implement of a control phase gate. With a $\sigma _{x}$ operation
on \textit{1'}, the nonlocal \textit{CNOT}\ has been demonstrated recently,
where qubit \textit{1 }and\textit{\ 1'} are path and polarization of photon 
\textit{A}\cite{HRYDG}. Moreover, it can be generalized to the controlled
rotation $CR=\left\vert 0\right\rangle _{1^{\prime }1^{\prime }}\left\langle
0\right\vert \otimes U_{com}+\left\vert 1\right\rangle _{1^{\prime
}1^{\prime }}\left\langle 1\right\vert \otimes U_{com}^{\prime }$ for any $%
U_{com}$ and $U_{com}^{\prime }$. In our experiment, \textit{1'} is the
wavepacket distinction and the mode-mismatch, both of which are traced out.
So both the rotation and the dephase on photon \textit{A} are teleported.
The dephased operation is a complete positive map ${\Large \varepsilon }_{d}%
{\Large =}\{\sqrt{\frac{1+p\eta }{2}}U_{com},\sqrt{\frac{1-p\eta }{2}}%
U_{com}\sigma _{z}\}$. The final state after operation is 
\begin{equation}
\rho _{d}(\left\vert \psi \right\rangle )=\varepsilon _{d}(\left\vert \psi
\right\rangle )=\left( 
\begin{array}{cc}
\alpha \alpha ^{\ast } & p\eta \alpha \beta ^{\ast }e^{-i2\varphi } \\ 
p\eta \alpha ^{\ast }\beta e^{i2\varphi } & \beta \beta ^{\ast }%
\end{array}%
\right) .
\end{equation}

To completely characterize the remote operation $\varepsilon _{e}$ in our
experiment, four state $\{\left\vert H\right\rangle ,\left\vert
V\right\rangle ,\left\vert D\right\rangle ,\left\vert R\right\rangle =\frac{1%
}{\sqrt{2}}\left( \left\vert H\right\rangle -i\left\vert V\right\rangle
\right) \}$ are input for the quantum process tomography\cite{NC}\cite{ABJ}.
The process is represented by a positive Hermitian matrix $\chi =\{\chi
_{mn}\}$, which satisfies $\varepsilon (\rho )=\sum_{mn}\chi _{mn}E_{m}\rho
E_{n}^{\dagger }$ ($\{E_{m}\}=\{I,\sigma _{x},\sigma _{y},\sigma _{z}\}$).
The matrix are shown in Fig.4 for ideal rotation $\chi _{i}$, dephased
rotation $\chi _{d}$, and effective operation $\chi _{e}$ in our experiment,
where two parameters for $\varepsilon _{d}$ are measured, $p\approx 0.85$
and $\eta \approx 0.92$. And the comparison of experimental operation $%
\varepsilon _{e}$ with the dephased one $\varepsilon _{d}$ is characterised
by the average fidelity of pure state inputted throughout the Bloch sphere $%
\overline{F}[\varepsilon ^{\prime },\varepsilon ]=\int d\psi F[\varepsilon
^{\prime }(\left\vert \psi \right\rangle ),\varepsilon _{d}(\left\vert \psi
\right\rangle )]$, where $F[\rho ,\rho ^{\prime }]=Tr[\sqrt{\sqrt{\rho
^{\prime }}\rho \sqrt{\rho ^{\prime }}}]$\cite{NC}\cite{BOSBJ}\cite{N}. From 
$\chi $ we get $\overline{F}[\varepsilon _{i},\varepsilon _{e}]=0.96$ and $%
\overline{F}[\varepsilon _{d},\varepsilon _{e}]=0.99$. Geometrically, the
rotation can also be characterized by the actual rotation angle $\varphi
=\phi ^{\prime }-\phi $. Here we use the angle deviation $\delta =\delta
(\rho _{i},\rho _{e})$, which is defined by the cross angle between Bloch
vector by ideal rotaion and the one we finally get, to characterize the
experimental operation. The maximal angle is $\delta _{\max }=7^{\circ }$,
that means $\varphi =120^{\circ }\pm 7^{\circ }$.

In conclusion, a remote rotation on qubits throught $120^{\circ }$ about $z$
axis is performed using shared entanglement and local operation without
rotate the target photons. And the dephase on photon \textit{A} is also be
teleported. The whole process is measured by quantum process tomography and
agrees with the theoretical prodiction. The scheme can be generalized to
remote implement a class of controlled rotation. Although only rotaion
commuting with $\sigma _{z}$ is used in our experiment, it's the same with
operations anti-commuting with $\sigma _{z}$.\cite{HPV}

The authors acknowledge Yun-Feng Huang and Xi-Feng Ren for technique help.
This work was supported by the Chinese National Fundamental Research Program
(2001CB309300), the NSF of China (10304017), the Innovation funds from
Chinese Academy of Sciences.

\begin{quote}
\bigskip

Figure 1. a. Geometrical interpretation of single qubit rotaion $U_{com}$, $%
\left\vert \psi ^{\prime }\right\rangle =U_{com}\left\vert \psi
\right\rangle $; b. Quamtum circuit for schem of remote rotaion on single
qubit, where qubit 1 and 2 are entangled and qubit 3 is the target to be
operated on. The whole process is devided into three steps (see text for
details).

Figure 2. Experimental setup to perform a remote rotation on a single
photon. \textit{P1}, \textit{P2}: polarization beam splitters; $%
H_{1},H_{u},H_{d}$:half wave plates; $Q_{1},Q_{u},Q_{d}$: quarter wave
plates; P:polarizer; PA:polarization analyzer; IF: interference filter; $%
D_{1},D_{2}$:single photon detectors.

Figure 3. Schematic drawing for the mode-mismatching in polarization beam
splitter, which induces a dephase.

Figure 4. $\chi $ matrices determined by a) ideal rotaion $\chi _{i}$, b)
dephased rotaion $\chi _{d}$, and c) experimental rotaion $\chi _{e}$ from
fig. 2. The real parts of $\chi $ are on the left while the imaginary ones
are on the right.
\end{quote}

\end{document}